\documentclass[prl,nofootinbib,twocolumn,amssymb,amsmath]{revtex4-1}
\usepackage{graphicx} 
\usepackage[utf8]{inputenc}
\usepackage{multirow}
\usepackage{subfigure}
\usepackage{color}
\usepackage{hyperref}

\begin{document}

\title{Eigenstate Thermalization in Systems with Spontaneously Broken Symmetry}

\author{Keith R. Fratus}
\author{Mark Srednicki}
\affiliation{Department of Physics, University of California, Santa Barbara, California 93106, USA}

\begin{abstract}

A strongly non-integrable system is expected to satisfy the eigenstate thermalization hypothesis, which states that the expectation value of an observable in an energy eigenstate is the same as the thermal value. This must be revised if the observable is an order parameter for a spontaneously broken symmetry, which has multiple thermal values. We propose that in this case the system is unstable towards forming nearby eigenstates which yield each of the allowed thermal values. We provide strong evidence for this from a numerical study of the 2D transverse-field quantum Ising model.

\end{abstract}
\maketitle

The eigenstate thermalization hypothesis (ETH)
can explain how an isolated, quantum many-body system in an initial pure state
can come to thermal equilibrium 
(as determined by measurements of a specified set of observables) 
in finite time \cite{deutsch,srednicki,rdo}.
ETH is expected to hold in systems without disorder 
that are sufficiently far (in parameter space) 
from points of integrability,
for observables that are sufficiently simple (e.g. local) functions of 
the fundamental degrees of freedom.
In recent years ETH has been the subject of intensive analytic and 
numerical investigations, e.g. \cite{rdo,rigol,sr,rs,kpsr,iwu,bmh,skngg,kk,svph,kih,gg};
see \cite{pssv} for an overview including the connection to experimental
results in cold atoms and other systems.

The key statement of ETH is that expectation values of a relevant
observable $M$ in an energy eigenstate $|\alpha\rangle$
(of the full many-body hamiltonian $H$) take the form
\begin{equation}
\langle\alpha|M|\alpha\rangle = {\cal M}(E_\alpha),
\label{M}
\end{equation}
where ${\cal M}(E)$ is a smooth function of $E$ and $E_\alpha$ is the energy eigenvalue. 
In a system with $N\gg 1$ degrees of freedom,
this is enough information to show that 
${\cal M}(E)$ is equal, up to $O(N^{-1/2})$ corrections, to the canonical
thermal average of the operator $M$,
\begin{equation}
{\cal M}(E) = {\mathop{\rm Tr} M e^{-H/kT}\over \mathop{\rm Tr} e^{-H/kT}}
\bigl[1 + O(N^{-1/2})\bigr],
\label{ME}
\end{equation}
where the temperature $T$ is implicitly determined as a function of energy $E$ by
the usual relation
\begin{equation}
E = {\mathop{\rm Tr} H e^{-H/kT}\over \mathop{\rm Tr} e^{-H/kT}}.
\label{E}
\end{equation}

A second key statement of ETH is that the off-diagonal matrix elements of $M$ in
the energy basis, $\langle\alpha|M|\beta\rangle$ with $\alpha\ne\beta$,
are exponentially small in $N$.  This is needed to explain why the diagonal
matrix elements of eq.~(\ref{M}) dominate the instantaneous expectation value
of $M$ (in a generic time-dependent state) at almost all times, 
which in turn is necessary for thermal equilibrium to be maintained
once it has been achieved. However this aspect of ETH will not be our focus.

The ETH paradigm must be revisited for a system that is capable of 
exhibiting spontaneous symmetry breaking (SSB). Suppose that the observable $M$ 
is an order parameter for a global symmetry.  At energies corresponding
to the broken-symmetry phase, and in the infinite-volume limit, we expect the system
to have states with the same energy but with different values of $M$ (that are related
by the symmetry). In this case, eq.~(\ref{M}) cannot hold as written. 
We conjecture that, instead, the single smooth function ${\cal M}(E)$
is replaced by a multivalued function, with one branch for each 
allowed value of the order parameter.

We note that the compatibility of ETH and SSB was assumed to hold in \cite{zkh}, in which the tunnelling dynamics of the order parameter were studied in ``Schrodinger cat'' states of a quantum Ising model with disordered infinite-range interactions. 
This model has some special features that were the main concern of \cite{zkh}, 
and so an investigation of the basic issue in simpler models is warranted. Additionally, we focus directly on the equilibrium values of the order parameter rather than the quantum dynamics
of selected states.

We therefore turn our attention to a well-known and much studied model, the quantum transverse-field Ising model with constant nearest-neighbor interactions, specified by the hamiltonian
\begin{equation}
\label{eqn:hamform}
H = - \sum_{\langle i j \rangle } \sigma_{i}^{z}\sigma_{j}^{z} - g \sum_{i} \sigma_{i}^{x} .
\end{equation}
Here $\sigma^z_i$ and $\sigma^x_i$ are the usual Pauli matrices at a lattice site $i$, 
and the first term is a nearest-neighbour sum over the links of the lattice. Due to the presence of the transverse field term, this is a fully interacting quantum system, and in more than one dimension it is nonintegrable. This hamiltonian is 
invariant under the $Z_2$ spin-flip transformation generated by the unitary operator
$X=\prod_i \sigma^x_i$: $XHX^{-1}=H$. In two dimensions,
in the infinite-volume limit, this model exhibits a quantum phase transition at a critical coupling
$g_{c} \simeq 3.044$ \cite{crit}. 
For $g>g_c$, the ground state $|0\rangle$
is unique and satisfies $X|0\rangle =|0\rangle$. The magnetization operator 
\begin{equation}
M = \sum_{i} \sigma_{i}^{z}
\label{Mop}
\end{equation}
is odd under the symmetry, $XMX^{-1}=-M$, and 
has zero ground-state expectation value, $\langle0|M|0\rangle = 0$.
For $g<g_c$, the ground state is two-fold degenerate, 
with $\langle0{\pm}|M|0{\pm}\rangle=\pm{\cal M}_0$. The two ground states are related
by the symmetry, $X|0{\pm}\rangle=|0{\mp}\rangle$.
At finite temperature with $g<g_c$, there are correspondingly two phases separated
by a second-order phase transition at a critical temperature $T_{c}$. For $T<T_c$,
the thermal expectation value of $M$ has two values $\pm{\cal M}(E)$, where $E$
is related to $T$ by eq. (\ref{E}). At higher temperatures, ${\cal M}(E)$ vanishes. 
A fluctuation-corrected mean-field computation of ${\cal M}(E)$ for infinite volume
can be extracted from the results of \cite{stratt}. 

At finite volume, more care is required. The energy eigenstates $|\alpha\rangle$ are discrete and expected to be nondegenerate (for $g\ne 0$). Each must then also be an eigenstate of $X$ with eigenvalue $\pm 1$. This implies that $\langle\alpha|M|\alpha\rangle$ must vanish, 
since $M$ is odd under $X$. However, for $g<g_c$ and at energies corresponding to $T<T_c$,
we expect the energy eigenstates to be unstable to a small 
symmetry-breaking perturbation.
We therefore modify the hamiltonian by adding $M$ with a coefficient $\epsilon$,
\begin{equation}
\label{eqn:opform}
H \to H + \epsilon  M,
\end{equation}
with $\epsilon\ll1$. This explicitly breaks the $Z_2$ symmetry by an amount that is small compared to the energy scales in $H$. 
As long as $\epsilon$ is not much smaller than the mean level spacing
(which itself is exponentially small in the volume for a large system),
we expect the exact energy eigenstates to be linear combinations of the unperturbed
eigenstates with (nearly) equal and opposite expectation values of $M$. 
In a thermodynamically large system, we expect the system to be unstable in this way to an infinitesimally small perturbation.

We investigate the validity of this picture by performing an exact diagonalization of $H$
on a $4\times 5$ lattice with periodic boundary conditions for various values of $g$
and with $\epsilon=10^{-3}$.
This is smaller (by about a factor of five) than the mean level spacing across the full spectrum, 
but is still large enough to mix nearby eigenstates.
In addition to the $Z_2$ spin-flip symmetry, there is a discrete translation symmetry (in each cartesian direction) and a parity symmetry. We present results on the zero-momentum, even parity sector, which has 14,676 states (to be compared with 1,048,576 states in the full Hilbert space). We find comparable results in other sectors and with other (small) values of $\epsilon$.
We compute $\langle\alpha|M|\alpha\rangle$ for each state, and compare with
the infinite-volume, fluctuation-corrected mean-field value ${\cal M}(E_\alpha)$ 
computed using the equilibrium methods of \cite{stratt}. 
We use the results for coordination number 
$z=6$ with a rescaled value for $g$ and $M$, since the direct results for $z=4$ exhibit unphysical features such as a first-order phase transition and nonconvexity of the critical curve;
this scaling method is exact in mean-field theory.
This calculation is an essentially uncontrolled approximation, but it provides a useful
benchmark for our numerical results for individual eigenstates at finite volume.

\begin{figure}[h]
\centering
\includegraphics[width=85mm]{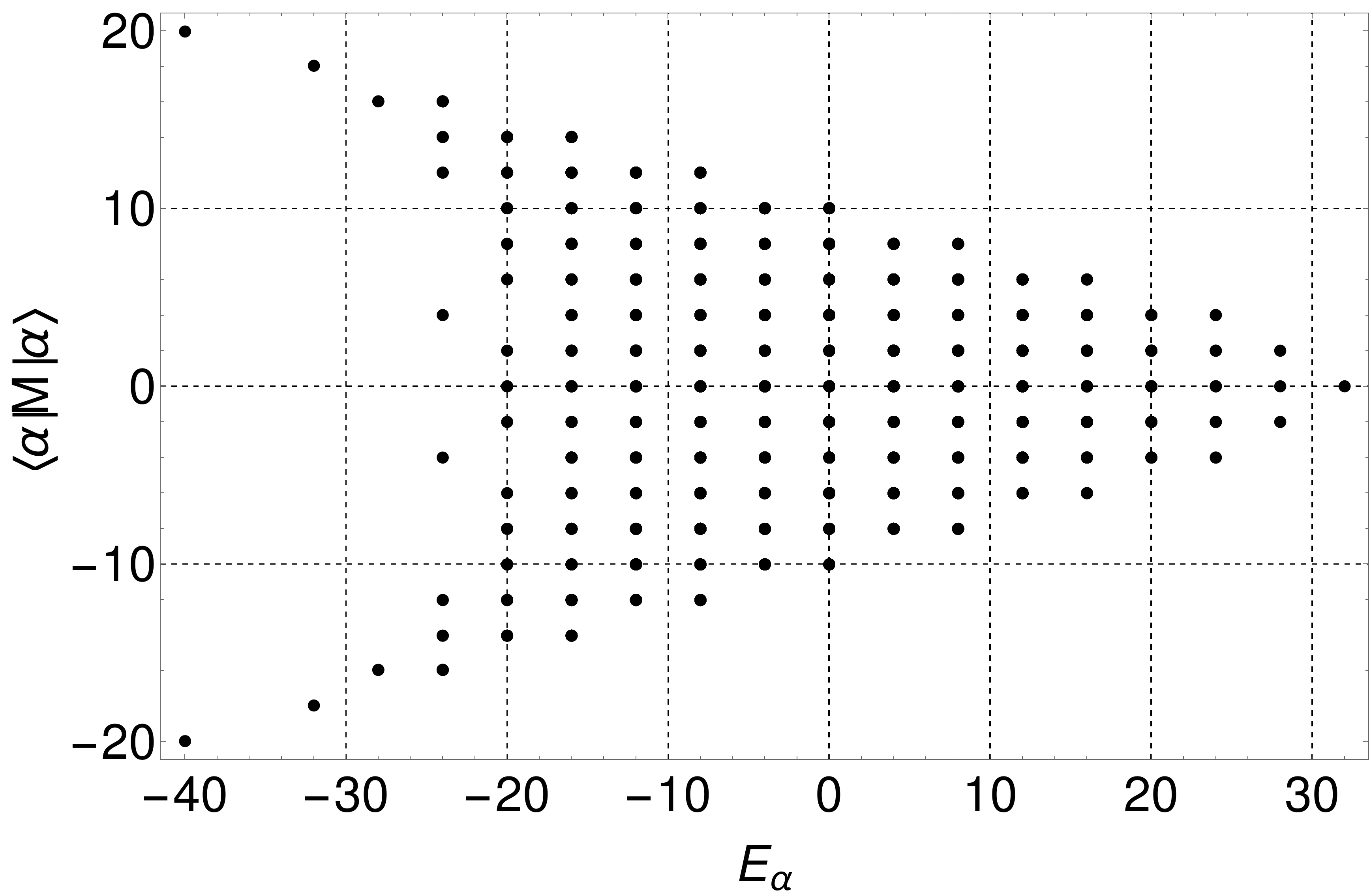}
\caption{The magnetization as a function of energy for energy eigenstates of
the 2D Ising model on a $4\times 5$ lattice with periodic boundary conditions and zero transverse field.}
\label{fig:plot0}
\end{figure}

As a point of comparison, we show results for the $g=0$ model in Fig. \ref{fig:plot0}.
In this case, the energy eigenvalues are integers ranging from $-40$ to $+32$. 
Each energy eigenstate is degenerate, and can be chosen to be a simultaneous
eigenstate of each $\sigma^z_i$; the magnetization $M$ is then 
obtained by summing these eigenvalues.
In this system, ETH is clearly not satisfied:
for every energy eigenvalue below the maximum, there are eigenstates with a range of values
of $M$. The large number of magnetization values for each energy can be understood from the fact that states with different net magnetization can result in the same nearest-neighbor bond energies, 
depending on how the individual spins are arranged into ``droplets'' of different sizes.

\begin{figure}[h]
\centering
\includegraphics[width=85mm]{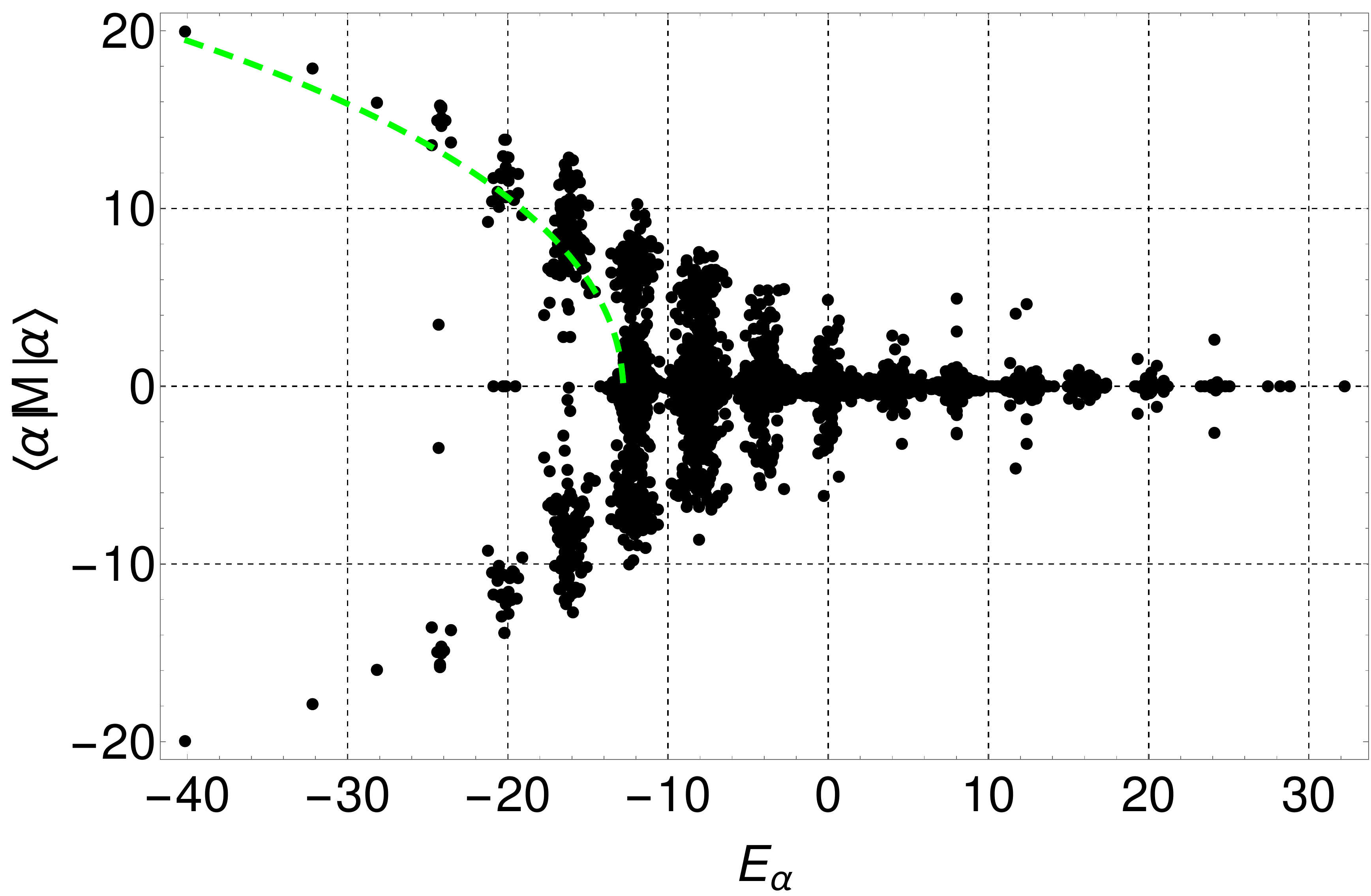}
\caption{Same as Fig. \ref{fig:plot0}, but with a transverse field with coefficient $g=0.25$. 
Only states with zero momentum and even parity are shown.
Dashed line: fluctuation corrected mean-field prediction in the thermodynamic limit, 
assuming ETH.}
\label{fig:plot25}
\end{figure}

Next we consider a small but finite value for the transverse field coefficient, $g = 0.25$.
Now we find that the energy eigenstates are all nondegenerate (even for $\epsilon = 0$), as expected.
We compute the expectation value of $M$ in each eigenstate in the zero-momentum,
even-parity sector, and show the results in Fig. \ref{fig:plot25}.
We see that for energy greater than roughly zero, the magnetization of every eigenstate has been compressed towards a value of zero. Below zero energy, the magnetization for each energy eigenstate has moved closer to one of two possible values, one positive and one negative.  However, these two branches are not sharply defined, 
indicating that ETH is not well satisfied for this value of $g$.
The dashed line shows the result of fluctuation-corrected 
mean-field calculation in the infinite volume limit, which is in fair agreement
with the numerical results.

\begin{figure}[h]
\centering
\includegraphics[width=85mm]{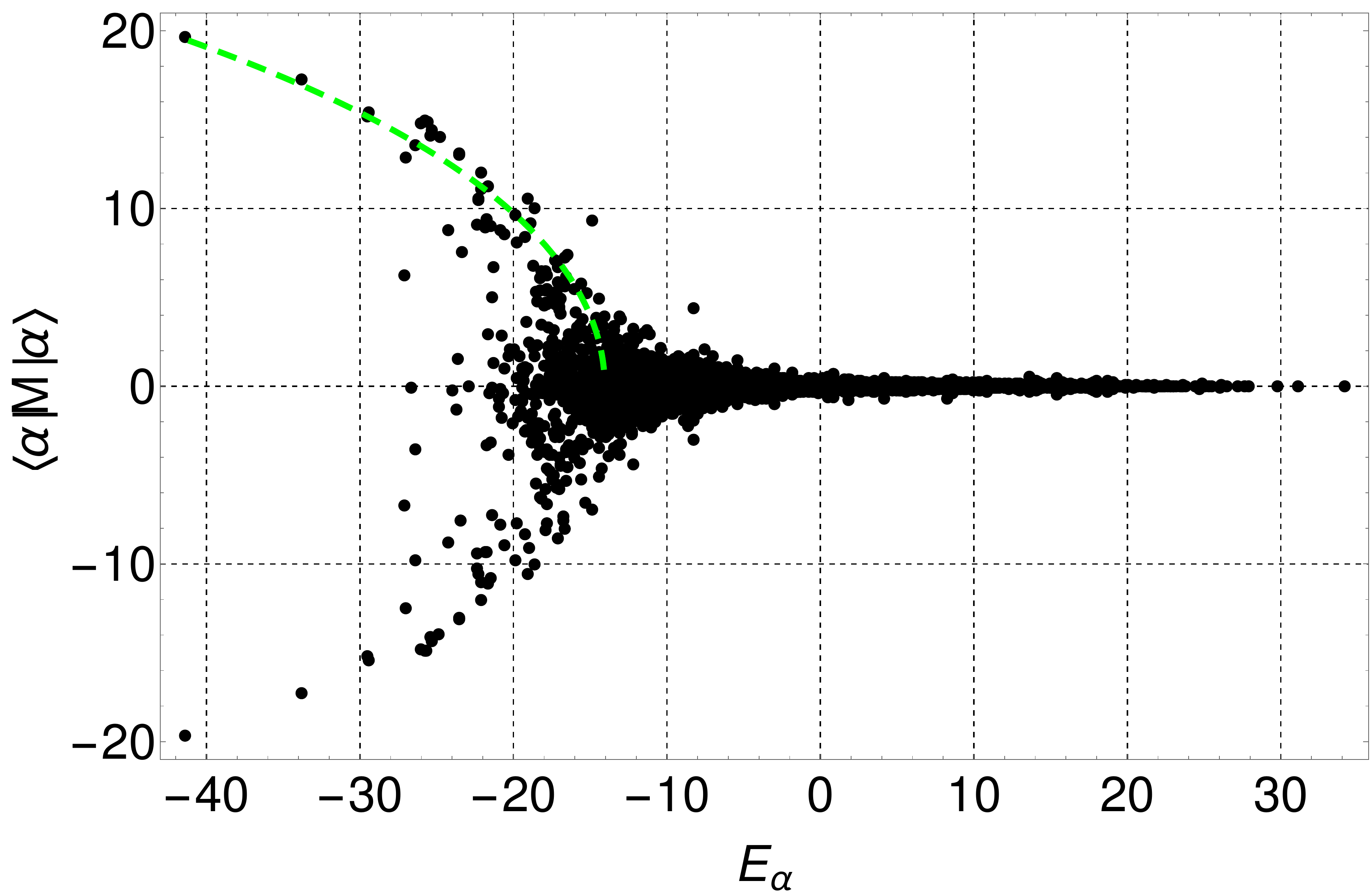}
\caption{Same as Fig. \ref{fig:plot25}, but with a transverse field with coefficient $g=0.75$.}
\label{fig:plot75}
\end{figure}

Results for $g = 0.75$ are shown in Fig. \ref{fig:plot75}.
There is now a clear qualitative difference between the energy eigenstates in the upper and lower portion of the spectrum. Above $E \simeq -5$, all states have a near-vanishing magnetization, while below $E \simeq -22$, almost all states possess a net magnetization which lies in one of two branches, one positive and one negative. 
Between these two energies, the separation between the two branches is less pronounced as they merge into a single line at zero magnetization. We see good agreement with the fluctuation-corrected mean-field calculation. 

\begin{figure}[h]
\centering
\includegraphics[width=85mm]{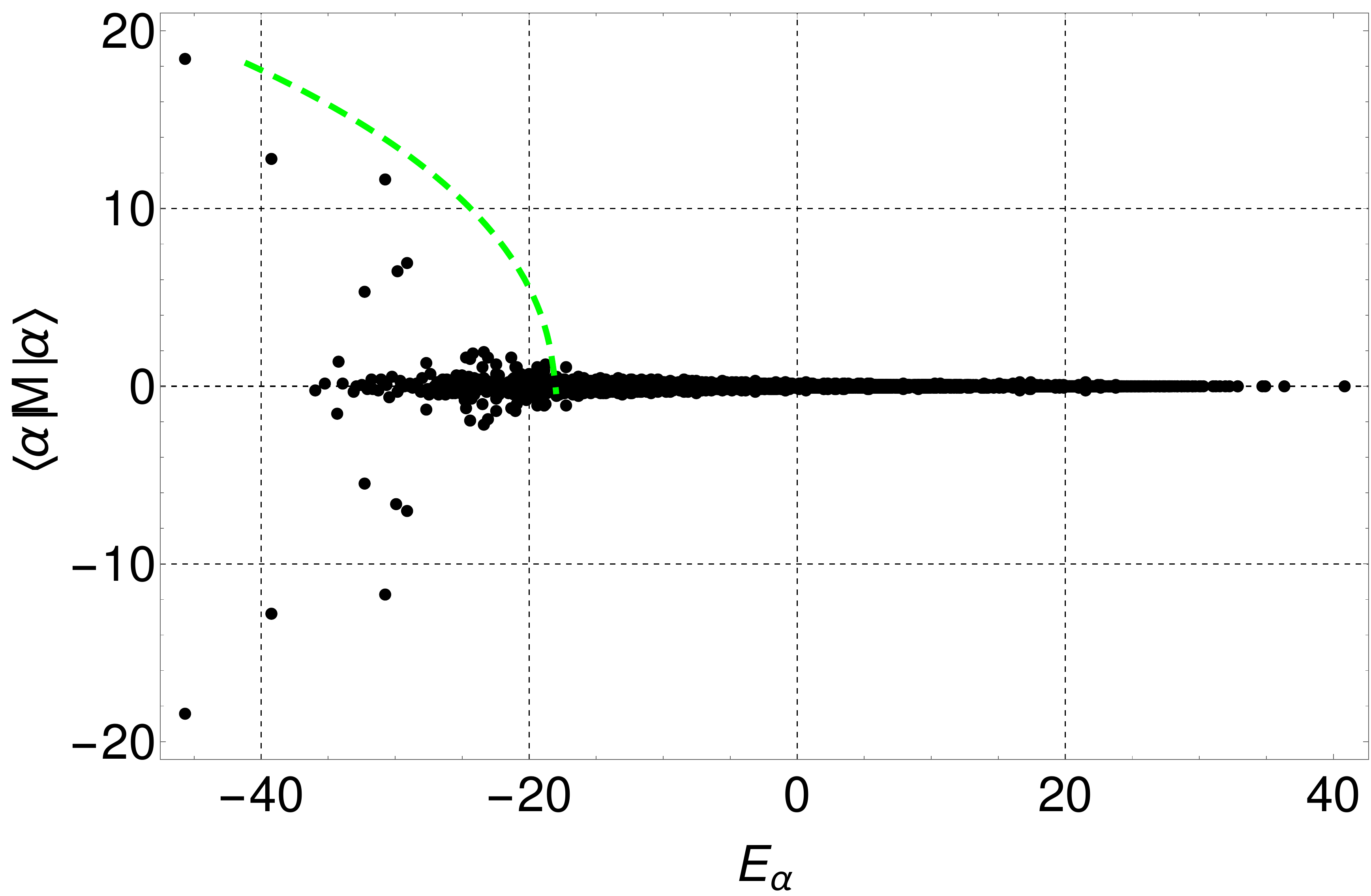}
\caption{Same as Fig. \ref{fig:plot75}, but with a transverse field with coefficient $g=1.5$.}
\label{fig:plot15}
\end{figure}

Results for $g = 1.5$ are shown in Fig. \ref{fig:plot15}.
Two magnetization branches can still be discerned, but
they are much less populated, due to the large transverse field lowering the critical energy
for spontaneous symmetry breaking. Rough agreement with the infinite-volume estimate
is still seen, though we do not expect our $z=6$ rescaling procedure be as accurate for this larger value of $g$.

\begin{figure}[h]
\centering
\includegraphics[width=85mm]{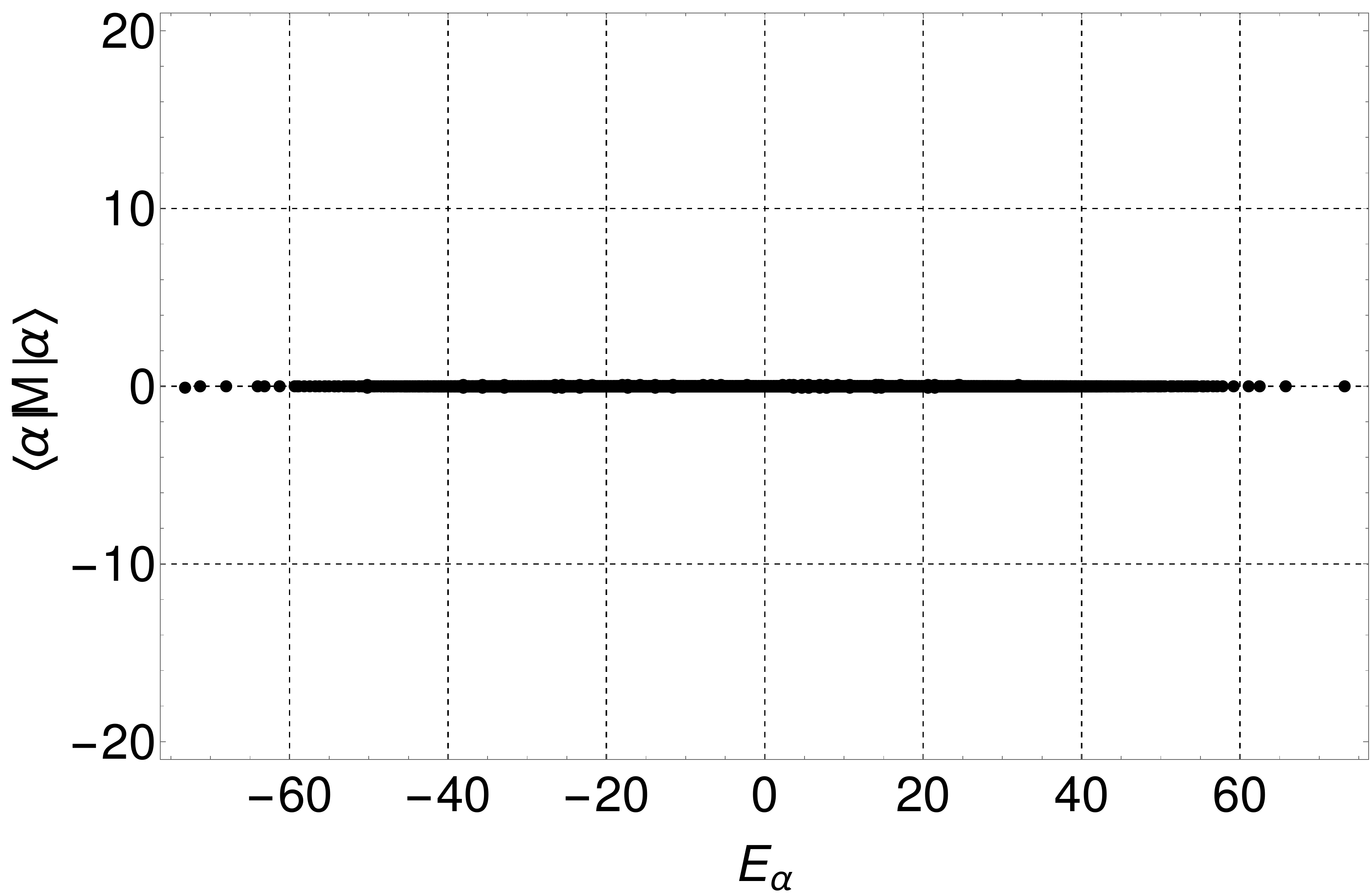}
\caption{Same as Fig. \ref{fig:plot75}, but with a transverse field with coefficient $g=3.5$.}\label{fig:plot35}
\end{figure}

Finally, we show results for $g=3.5$ in Fig. \ref{fig:plot35}. 
Now we expect to be in the unbroken phase at all energies, 
and we indeed find zero magnetization for all eigenstates.

In a much larger system, with values of $g$
large enough to be sufficiently far from integrability but less than the critical value $g_c$,
we expect that the magnetization branches would be much better defined, with a spread in values that is controlled by the symmetry-breaking coefficient $\epsilon$,
as long as $\epsilon$ is above a minimum value that is exponentially small
in the number of sites.

We also note that at a sufficiently large positive energy, corresponding to a
negative temperature, we expect a second phase transition from a disordered phase
to an antiferromagnetically ordered phase. An order parameter for this phase transition
is a staggered magnetization, Eq.~\eqref{Mop} with a minus sign inserted in the sum
on alternating sites. We have not attempted to study this transition.

We conclude that our numerical results are fully consistent with the coexistence of spontaneous symmetry breaking and eigenstate thermalization. At system energies where spontaneous
symmetry breaking can occur, for an observable $M$ that functions as an 
order parameter, the single smooth function ${\cal M}(E)$ of
eq. (\ref{M}) must be replaced by a multivalued function. 
After a small symmetry-breaking perturbation is turned
on, the expectation value of $M$ in an individual
energy eigenstate $\langle\alpha|M|\alpha\rangle$ will lie on one of these branches,
with nearby (in energy) eigenstates yielding expectation values on the other branches.

Systems that exhibit eigenstate thermalization have many key physical properties encoded in a single eigenstate, including some nonlocal properties \cite{gg}. Since, as we have seen, it is possible to accommodate spontaneous
symmetry breaking within the ETH paradigm, this should extend to critical phenomena at energies near a second-order phase transition. Thus it should be possible, in principle, to extract critical exponents from a single eigenstate. Study of this question is currently limited by small system sizes, but we hope to return to it in future work.

\begin{acknowledgements}

We thank James Garrison, Tarun Grover, David Huse, and Marcos Rigol for helpful discussions,
and Sebastian Fischetti for computer support.
This work was supported in part by NSF Grant PHY13-16748.
We acknowledge support from the Center for Scientific Computing from the CNSI, MRL: an NSF MRSEC (DMR-1121053) and NSF CNS-0960316. 

\end{acknowledgements}

\end{document}